\documentclass[aip,amsmath,pre,amsfonts,
preprint,
author-numerical]{revtex4-2}
\usepackage{xcolor}
\usepackage{hyperref}
\usepackage{graphicx}
\usepackage{caption}
\usepackage{subcaption}
\usepackage{amssymb}

\begin{document}
	\title{Hard-core attractive Yukawa fluid global isomorphism with the lattice gas model}
	\author{A.~Katts}
	\email{a.katts@stud.onu.edu.ua}
	\author{V.L. Kulinskii}
	\email{kulinskij@onu.edu.ua}
	\affiliation{Department for Theoretical
		Physics and Astronomy, Odessa National University, Dvoryanskaya 2, 65082 Odessa, Ukraine}
	\begin{abstract}
		In this work we study the global isomorphism between the liquid-vapor equilibrium of the hard-core attractive Yukawa fluid (HCAYF) and that of the Lattice Gas (LG) model of the Ising-like type. The applicability of the global isomorphism transformation and dependence of its parameters on the screening length of the Yukawa potential are discussed. These parameters determine both the slope of the rectilinear diameter of the liquid-vapor binodal and the Zeno-element which are the core ingredients of the fluid - lattice gas isomorphism. We compare the Zeno-element parameters with the virial Zeno-line parameters which are commonly used in the literature for the formulation of generalized law of the correspondent states. It is demonstrated that the Zeno-element parameters appear to be sensitive to the liquid state instability when the interaction potential becomes too short-ranged while the virial ones do not show any peculiarities connected with this specific of the HCAYF.
	\end{abstract}
	\maketitle
	\section{Introduction}\label{sec:intro}
	The liquid-vapor equilibrium of the majority of atomic and molecular fluids shows common regularities. Probably the oldest one is the law of the rectilinear diameter (LRD) \cite{crit_diam1}: 
	\begin{equation}
	\rho_{d} = \frac{\rho_{l}+\rho_{g}}{2\,\rho_c}= 1+A\,\left(\, 1-T/T_c \,\right)\, .
	\label{eq:rdl}
	\end{equation}
	Another regularity is the approximate linearity of unit compressibility factor \cite{eos_zenoholleran_jcp1967,eos_zenonedostup_1979,eos_zenoboyle_jphysc1983,eos_zenobenamotz_isrchemphysj1990}:
	\begin{equation}\label{eq:z1}
	Z = \frac{P}{\rho\,T} = 1\,,
	\end{equation}
	where $\rho$ is the density, $P, T$ are the pressure and the temperature correspondingly. Eq.~\eqref{eq:z1} defines a line in $\rho - T$ plane which is called the Zeno-line \cite{eos_zenobenamotz_isrchemphysj1990}. For the case of van der Waals equation of state (EoS):
	\begin{equation}
	p=\frac{\rho\,T}{1-\rho\,b}-a\,\rho^2
	\end{equation}
	the line Eq.~\eqref{eq:z1} is exactly straight (the Batchinsky's law) \cite{eos_zenobatschinski_annphys1906}:
	\begin{eqnarray}\label{eq:z1vdw}
	\frac{T}{T^{(vdW)}_{B}}+\frac{\rho}{\rho^{(vdW)}_B} =& 1\\
	T^{(vdW)}_{B} =a/b \,, \quad \rho^{(vdW)}_{B} =&\, 1/b \notag
	\end{eqnarray}
	Experimental data reveal that approximate linearity of Eq.~\eqref{eq:z1} takes place for a vast set of molecular fluids \cite{eos_zenoboyle_jphysc1983,eos_zenoapfelbaum1_jpcb2009,eos_zenosanchez_jpcb2016}. Model equation of states which are widely used in chemical thermodynamics such as Peng-Robinson and Redlich-Kwon can reproduce the linearities Eqs.~\eqref{eq:rdl} and \eqref{eq:z1} only at a cost of parameter fitting  \cite{eos_zeno_jphyschemb2000,eos_eosdiamzeno_intjthermodyn2004,eos_zenoherschbachktrans_jpcb2013}. Sure this is important for applications and the description of real fluids. But such an approach does not provide physical insight as to the nature of these regularities.
	In terms of the standard approach based on the closure for the Ornstein-Zernike integral equation the question reduces to the relevant  approximation for bridge-functional \cite{eos_zenoline_jphyschem1992,eos_zenoidealinesarkisov_jcp2002}. In this respect numerical methods for calculating the grand canonical-function are of great importance because they allow to trace the dependence of thermodynamic quantities on the parameters of various interaction potentials \cite{liq_zenosupercrit_cpl2016,eos_zenodesgrangeswater_jcp2016}. The results show that equation  \eqref{eq:z1} valid even for ionic liquids and liquid metals \cite{eos_zenoapf_morseiron_jcp2011,eos_apfelberill_jpcb2012,eos_zenometals_jpcb2015,eos_zenoliqmetalscritscal_jpcb2016,eos_zenolineionic_cpl2017}.
		
	The abundance of linearities Eqs.~\eqref{eq:rdl} and \eqref{eq:z1} among different fluid systems gives a hint that some extended principle of corresponding states can be formulated for a wide class of potentials based on some general concept. We believe that the idea of the liquid-gas state triangle   \cite{eos_zenoapfelbaum_jpchema2004,eos_zenotriapfelbaum_jpchemb2006} serves as promised ground. On this basis the concept of global isomorphism between a fluid and a lattice gas was proposed in \cite{eos_zenome0_jphyschemb2010,eos_zenomeglobal_jcp2010}.
	
	Obviously, in the limit $\rho\to 0$ the Zeno-line can be obtained by the virial expansion \cite{eos_zenonedostup_1979,eos_zenobenamotz_isrchemphysj1990,eos_zeno_jphyschem1992}: 
	\begin{equation}\label{eq:zenovir}
	\frac{\rho}{\rho_B} + \frac{T}{T_B} = 1
	\end{equation}
	where the parameters $T_{B}$ and $\rho_{B}$ is determined by the virial coefficients:
	\begin{equation}\label{eq:tbnb}
	B_2(T_{B}) = 0\,,\quad \rho_{B}= \frac{T_B}{B_3\left(\,T_B\,\right)}\,\left. \frac{dB_2}{dT}\right|_{T= T_B}\,.
	\end{equation}
	The straight line \eqref{eq:zenovir} is called the virial Zeno-line \cite{eos_zenoline_potentials_jcp2009}. 
	
	Fluids with hard core Yukawa interaction of the form:
	\begin{equation}\label{eq:yukwpot}
	\Phi(r) =\begin{cases}
	\infty\,, & \text{if}\quad r< \sigma \\
	-\frac{\varepsilon}{r/\sigma} \exp\left(\,-\lambda (r/\sigma -1) \,\right)\,, & \text{if} \quad r\ge \sigma\,,
	\end{cases}
	\end{equation}
	are important class of model systems where \eqref{eq:rdl} and \eqref{eq:z1} are observed. The parameter $\lambda$ determines the range of attractive interaction and $\sigma$  is the hard core radius of a particle. The potential \eqref{eq:yukwpot} is widely used for the description of systems with screened coulombic interactions like dusty plasma \cite{thermodyn_dustyplasma_ufn2004ru}, associated fluids,  globular protein solutions \cite{thermodyn_yukawaproteins_jpcm2004}, fullerens etc. \cite{eos_ljsimplefluids_jcp2006,eos_fullerens_jcp2009}.  According to results of \cite{eos_yukawacorrstates_jcp2008} the LRD and the law of corresponding states for the HCAYF are observed for $\lambda\lesssim 7$ where the liquid state is considered stable. So one may expect that the Zeno-line is straight also at least for $\lambda<4$ when the stability region is great enough \cite{eos_zenoline_potentials_jcp2009}. 
	Within the concept of triangle of liquid-gas states   \cite{eos_zenoapfelbaum_jpchema2004,eos_zenotriapfelbaum_jpchemb2006} the disappearance of the liquid phase stability can be viewed as the degeneration of such triangle because of the singular behavior of the Zeno-line parameter $\rho_B$. 
	We note here that the construction of the triangle of liquid-gas states uses the supposition that the virial Zeno-line is the tangent to the alleged continuation of the binodal into low temperature region. But it is clear that such a procedure is quite ambiguous. Such an ambiguity can be circumvented in the fluid-lattice gas global isomorphism (GI) approach \cite{eos_zenome0_jphyschemb2010,crit_globalisome_jcp2010}. The main idea is the topological similarity between the phase diagram of the fluid and that of Ising model (lattice gas). The mapping between the liquid-vapor equilibrium curve and the lattice gas binodal was constructed (see also \cite{eos_zenoapfvorob_lattice2real_jpcb2010}). By this virtue the thermodynamic characteristics of a fluid can be calculated on the basis of corresponding ones of the lattice model. The analytical form of such GI mapping is determined by the LRD \eqref{eq:rdl}. In such a case it has simple form of the projective transformation of $\rho-T$ plane. The linearity of the Zeno-line \eqref{eq:z1} plays minor role as it is assumed that the binodal is inscribed into the triangle where the diameter is a median. The parameters of such triangle in general do not match with \eqref{eq:tbnb}. Rather their values are based on the Zeno-element \cite{eos_zenomeunified_jphyschemb2011} which is identified with the extreme state $x=1$ of the isomorphic lattice model where all cites are occupied. This approach was successfully used to relate the critical compressibility factor $Z_c$ of the Lennard-Jones fluids with the critical value of the Gibbs potential of the Ising model  \cite{eos_zenozcme_jcp2013,eos_zenozcassocme_jcp2014}.
	The application of the GI idea to the Lennard-Jones fluid is based on the homogeneity property of the attractive part of the potential. This is no more true for the HCAYF potential  \eqref{eq:yukwpot}. Nevertheless, the LRD \eqref{eq:rdl} and \eqref{eq:zenovir} also take place such class of fluids at least for intermediate interaction range with $0< \lambda\lesssim 7$   \cite{eos_yukawaduda_jcp2007,eos_zenoline_potentials_jcp2009}. Therefore it is important to test the applicability of the GI to the HCAYF and clarify the difference between parameters \eqref{eq:zenovir} and those of GI transformation.
	
	The aim of this work is to apply the global isomorphism concept to the liquid-vapor equilibrium of the HCAYF. We study the dependence of the parameters on the inverse screening length $\lambda$ and compare their behavior with  the virial ones Eq.~\eqref{eq:zenovir}. The paper is organized as follows. In Section~\ref{sec:globiso} we apply the global isomorphism transformation to symmetrize the data on the liquid-vapor equilibrium of the HCAYF and map them onto the binodal of the Ising model. In Section~\ref{sec:zenoelmntparam} the dependence of the isomorphism transformation parameters on $\lambda$ is investigated with special focus on the instability of liquid phase  in the short ranged limit $\lambda \gg 1$. We show that the Zeno-element parameters can be used to mark the liquid phase instability in contrast to the commonly used virial parameters of the Zeno-line \eqref{eq:zenovir}. Obtained results are compared with known analytical and numerical data for the value $\lambda_{*}$ at which the liquid branch disappears. The results are discussed in concluding section.
	
	\section{Global isomorphism transformation and binodal symmetrization}\label{sec:globiso}
	In \cite{eos_zenome0_jphyschemb2010} the mapping between the fluid and the lattice gas was proposed. It is based on the idea of the triangle of liquid-gas  states \cite{eos_zenotriapfelbaum_jpchemb2006} and topological similarity between corresponding part of diagrams of these systems. The Hamiltonian of the LG with the Ising-like site-site interaction is:
	\begin{equation}\label{ham_latticegas}
	H = -J\,\sum\limits_{
		\left\langle\, ij \,\right\rangle
	} \, n_{i}\,n_{j} - \mu\,\sum\limits_{i}\,n_{i}
	\end{equation}
	Here $J$ is the interaction energy between nearest neighbors, $n_{i}$ is the site filling number and $n_{i} = 0,1$ whether the site is empty or occupied correspondingly and $\mu$ is the chemical potential \cite{book_baxterexact}.
	
	Assuming the validity of the LRD \eqref{eq:rdl} such a mapping has the form of projective transformation:
	\begin{equation}\label{eq:projtransfr}
	\rho =\, \rho_{*}\,\frac{x}{1+z \,\tilde{t}}\,,\quad
	T =\, T_{*}\,\frac{z\, \tilde{t}}{1+z \,\tilde{t}}\,,\quad z = \frac{T_c}{T_{*} - T_c}
	\end{equation}
	Corresponding diameter is:
	\begin{equation}\label{eq:zenodiam}
	\frac{\rho_{l}+\rho_{g}}{2\,\rho_c} = 1 + z\,\left(T/T_c-1\right) 
	\end{equation}
	Here $x=\left\langle n_{i}\right\rangle$, $ \tilde{t}=t/t_{c}$ is the cite occupation probability and the temperature of the LG reduced to its critical value $t_{c}$ correspondingly. The temperature parameter $T_{*}$ is determined by the Boyle point in the van der Waals approximation:   
	\begin{equation}\label{eq:tbvdwmy}
	B^{vdW}_2(T_{*}) = 0\,,\quad  T_{*} =  T^{(vdW)}_{B}  = \frac{a}{b}\,,
	\end{equation}
	where
	\begin{equation} \label{eq:vdw_ab}
	a =\,\, -2\pi\,\int\limits_{\sigma}^{+\infty}\Phi_{\text{attr}}(r)\,r^2\,dr\,.    
	\end{equation}
	Here $\Phi_{attr}(r)$ is the attractive part of the interaction potential $\Phi(r)$, $\sigma$ - particle diameter so that $b = \frac{2\pi}{3}\,\sigma^{3}$. Further common dimensionless definition for the temperature $T\to T/\varepsilon$ and the density $\rho\to \rho\,\sigma^{3}$ will be used throughout the paper. The Zeno-element density parameter $\rho_{*}$ is as following: 
	\begin{equation}\label{eq:nbvdwmy}
	\rho_{*}= \frac{ T_{*} }{B_3\left(\,T_{*}\,\right)}\,\left. \frac{d\,B_2}{dT}\right|_{T= T_{*}}\,.
	\end{equation}
	and corresponds to the LG state $x=1$ with all sites occupied. The parameters $\rho_{*},\, T_{*}$ determine the linear Zeno-element:
	\begin{equation}
	\frac{T}{T_{*}}+\frac{\rho}{\rho_{*}} = 1
	\label{eq:z1globiso}
	\end{equation} 
	%
	The physical reasoning behind Eqs. \eqref{eq:tbvdwmy} and  \eqref{eq:nbvdwmy} is as following. For potentials which are represented as a sum of attractive and repulsive parts the Hamiltonian of the isomorphic LG model has the attraction part only. Therefore the parameter $\rho_{*}$ in Eqs. \eqref{eq:tbvdwmy} and \eqref{eq:vdw_ab} is determined by this part of the fluid potential. The number of cites $\mathcal{N}$ of the isomorphic LG model is determined by the van der Waals approximation for fluid EoS as $\mathcal{N} = \rho_{*}\, V$ \cite{eos_zenomeunified_jphyschemb2011}. The Zeno-element is the image of the line $x=1$ for the LG. So the Zeno-element becomes the tangent to the liquid branch of the binodal at $T\to 0$. Such a definition of the limiting state $\rho_{*}$ circumvents the problem of ambiguity of continuation of the binodal beyond the triple point. Since the number of sites for two dual states $x\to 0$ and $x\to 1$ is the same it is clear how the properties at $\rho\to 0$ determines the liquid branch of the binodal.
	
	Despite similarity between the ``thermodynamic`` virial Zeno-line \eqref{eq:zenovir} and the ``geometric`` Zeno-element \eqref{eq:z1globiso} they differ in general. We illustrate this taking the LG model as an example. The mean-field approximation for EoS for LG is as following (see e.g. \cite{book_statmechdorlas}):
	\begin{equation}\label{eq:lgeos}
	P_{\text{LG}}(x,t) = -t\,\text{Ln}(1-x) - \left.x^2\right/2
	\end{equation}
	where $P_{\text{LG}}$ is the pressure and we use $\epsilon$ as the natural energy unit for the temperature variable $t$. Equation \eqref{eq:lgeos} is very much alike vdW EoS though entropy excluded volume term is in form of simple ideal gas approximation, i.e. particles have no hard core and interact
	via attractive potential only. Irrespective of the approximation for the EoS LG it has symmetrical binodal which reflects underlying particle-hole duality. The virial expansion for
	EoS \eqref{eq:lgeos}
	%
	determines the VZL on the basis of the Boyle point:
	\begin{equation}\label{eq:virialgzeno}
	\frac{t}{t_B}+\frac{x}{x_B}=1
	\end{equation}
	with $t_B=1, x_B=3/2$ as the virial Zeno - parameters according to \eqref{eq:tbnb}. Clearly, the value $x_{B}$ lies beyond physical region
	$0\le x \le 1$.
	\begin{figure}
		\includegraphics[width=0.8\linewidth]{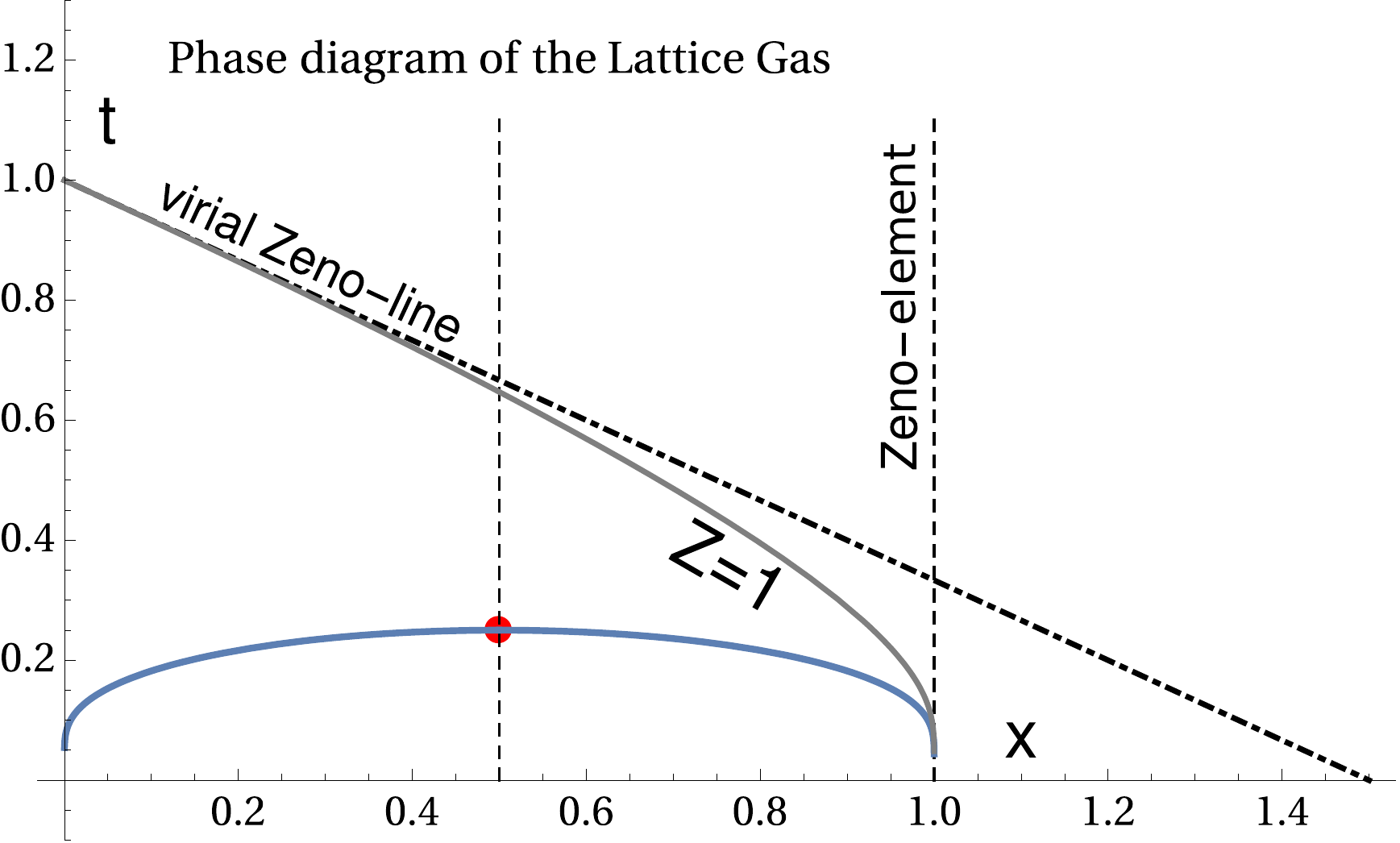}
		\caption{Zeno-states of the Lattice Gas EoS \eqref{eq:lgeos}. Solid gray line - Zeno-states \eqref{eq:lgzeno}, dot$\&$dashed - the virial Zeno-line
			\eqref{eq:virialgzeno}.}
		\label{fig_zlinelattgas}
	\end{figure}
	The exact set of Zeno-states for \eqref{eq:lgeos} can be found easily:
	\begin{equation}\label{eq:lgzeno}
	Z = \frac{P_{LG}}{x\,t}=1 \rightarrow  t_Z(x) = -\frac{x^2}{2(x+\text{Ln} (1-x))}
	\end{equation}
	and clearly is not linear (see Fig.~\ref{fig_zlinelattgas}). It
	is indeed tangent to the binodal at $t\to 0$ because of the asymptotic low-temperature behavior of the lattice model in this limit \cite{eos_zenomeglobal_springer2015,eos_zenocrystapfelbaumvorob_jpcb2020}. 
	Thus the ZL $Z=1$, the VZL \eqref{eq:zenovir} and the tangent to the low-temperature fluid branch of the binodal are different in general.  This example demonstrates that it is not the virial expansion of the EoS which represent the internal symmetry adequately and the Zeno-element should be used instead.
	It is clear that the Zeno-element, which in this case is the line $x=1$ reflects the intrinsic symmetry of the system, while thermodynamic Zeno-states \eqref{eq:lgzeno} and the VZL \eqref{eq:virialgzeno} are determined by the specific choice of state variables $x,t$. In addition, the absence of the hard core diameter makes the Boyle point with $t_{B} = 1$ in this case rather formal since there is no balance between attraction and repulsion in LG. 	The lattice spacing is free scale parameter and the diameter $\sigma$ of a ``particle`` equals zero and the fact that $t_{*}\to \infty$ with corresponding Zeno-element as the tangent to the binodal at $T\to 0$ represents such situation more adequately. 
	The fact that the Zeno-element parameters $T_{*},\rho_{*}$ should be used for the construction of mapping has been demonstrated for the case of the LJ fluid in 2 and 3 dimensions in \cite{eos_zenomeglobal_jcp2010,crit_globalisome_jcp2010, eos_zenozcme_jcp2013}.
	
	Though these results obtained within the mean-field approximation one can expect that they remain unchanged in essential by the critical fluctuations. Indeed, the Zeno-states lie well beyond the critical region where long-range fluctuations break down mean-field description.
	That is why we use the Zeno-element \eqref{eq:z1globiso}  defined on the basis of the vdW EoS as the universal approximation corresponding to the virial expansion of EoS \cite{book_ll5_en}. It serves as the basis for the mapping between the triangle of liquid-gas state and the phase diagram of the Ising model which are equivalent topologically. We will demonstrate the difference between the virial Zeno-line parameters $T_{B}, \rho_{B}$ and $T_{*}, \rho_{*}$ for HCAYF below.
	
	The transformation  \eqref{eq:projtransfr} restores the particle-hole symmetry which is inherent for the LG and allows to continue the binodal beyond the triple point using the LG binodal. The binodal symmetrization method along with the mapping onto its  lattice gas image was used in \cite{crit_globalisome_jcp2010} for the Lennard-Jones fluid in 2D and 3D cases (see also \cite{eos_zenoapfvorob_lattice2real_jpcb2010,eos_zenoherschbachktrans_jpcb2013,eos_zenovorobsymm_cpl2014,eos_zenometals_jpcb2015}). Here we use this method for the HCAYF in 3D. The computer simulation data \cite{eos_yukawacorrstates_jcp2008} for the HCAYF liquid-vapor equilibrium with $1\lesssim \lambda \lesssim 7$ are presented. 
	\begin{figure*}
		\includegraphics[width=0.95\textwidth]{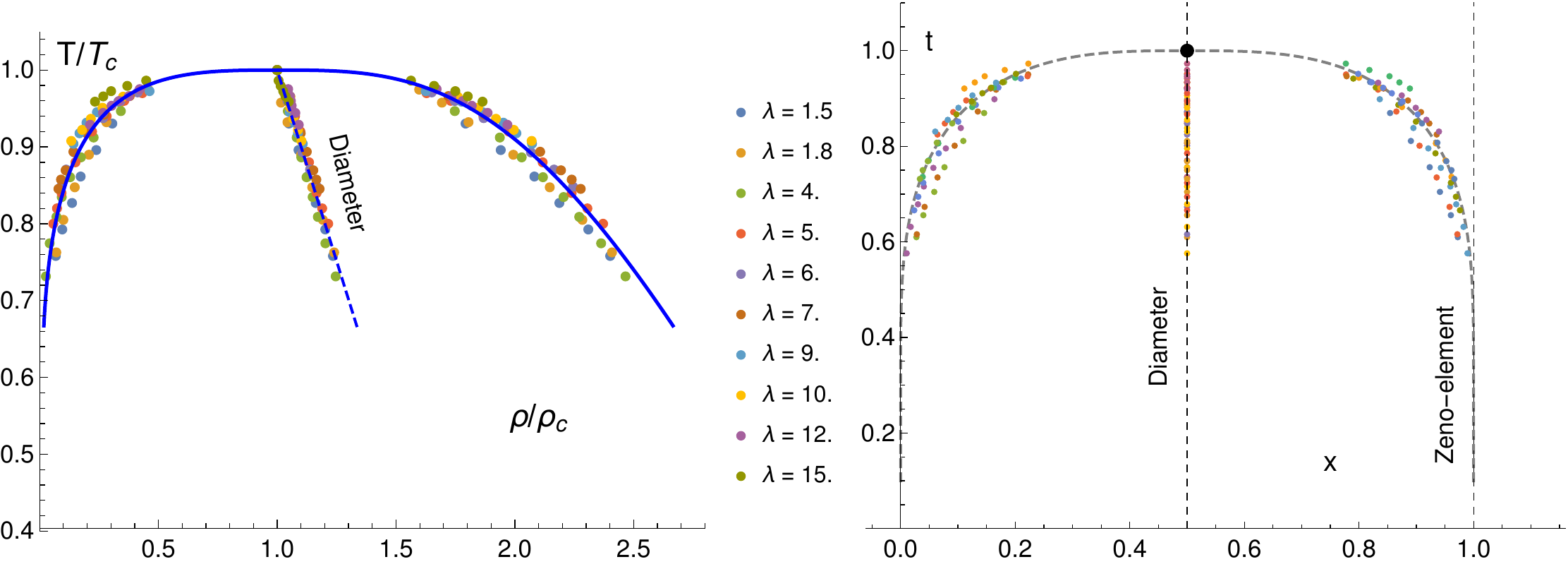}
		\caption{Symmetrization of the HCAYF binodal for $1.5\lesssim \lambda \lesssim 7$ according to the data of \cite{eos_yukawacorrstates_jcp2008}. Solid curve on the left plot - is image of $3D$ Ising model binodal obtained numerically in \cite{crit_3disingmc_jmathphys1996} mapped by \eqref{eq:projtransfr}  (see the text below).}
		\label{fig:symmbinod}
	\end{figure*}
	They clearly demonstrate the corresponding state principle in reduced coordinates $(\rho/\rho_c(\lambda),T/T_c(\lambda))$. Evidently, the transformation parameter $z$ depends on $\lambda$ but we do not discuss this dependence here and get simple estimate $z \approx 1$ consistent with the data. 
	
	Now we can check the isomorphism with the Ising model as obvious choice. According to \cite{crit_3disingmc_jmathphys1996} the Ising model binodal data  in the interval $5\cdot10^{-4}\le \tau\le 0.26\,,\,\,\tau=1-\tilde{t}$  can be fitted as: 
	\begin{equation}\label{eq:m3distalapovblote}
	x(t) =\frac{1}{2}\left(1 \pm \tau^{\beta} (b_0 - b_1\,\tau - 
	b_2*\tau^{\Delta})\right)\\
	\end{equation}
	with
	\[\beta = 0.327\,,\quad b_0\approx 1.69\,,\quad b_1 \approx 0.426\,,\quad b_2\approx 0.344\,,\quad \Delta \approx 0.51\]
	We map the binodal \eqref{eq:m3distalapovblote} onto the one of HCAYF \eqref{eq:projtransfr} using global isomorphism projective transformation \eqref{eq:projtransfr}. As one can see from Fig.~\ref{fig:symmbinod} the result is consistent with the statement about isomorphism between the lattice gas and  the HCAYF for $1\lesssim \lambda \lesssim 7$ with $z(\lambda)\gtrsim 1$. For 3D LJ fluids $z=1/2$ \cite{eos_zenomegenpcs_jcp2010,crit_globalisome_jcp2010}.
	\section{Zeno-element parameters of HCAYF}\label{sec:zenoelmntparam}
	In this Section we study the dependence of the parameters \eqref{eq:tbvdwmy},\eqref{eq:nbvdwmy} of the Zeno-element and compare it with that of the virial ones \eqref{eq:tbnb}. In accordance with the GI we search for the value of the screening parameter $\lambda_{*}$ which marks the disappearance of the liquid branch of the binodal.
	At first let us check the necessary condition $T_c<T_{*}$ for applying Eq.~\eqref{eq:projtransfr} which means that there is the isomorphic lattice model of the Ising type. As has been noted above for too short-ranged Yukawa potential the liquid phase becomes unstable \cite{eos_yukawafluid_jcp1994}. 
	Once $T_{*}$ or $T_B$ approaches $T_{c}$ the liquid branch becomes shorter the triangle of liquid-gas states degenerates as the binodal tangent at the triple point and goes almost horizontally. This means that the corresponding density parameter should tend to infinity. In physical terms the limiting density either $\rho_{B}$ or $\rho_{*}$ corresponds to the dense ``liquid`` state with all sites filled. Therefore if  the potential is very short-ranged such a state  can be stable only at very high density and one should expect the singular behavior of the density parameters $\rho_{B}$ and $\rho_{*}$ representing the instability of the liquid phase. 
	
\begin{figure}
		\centering
		\includegraphics[width=0.7\linewidth]{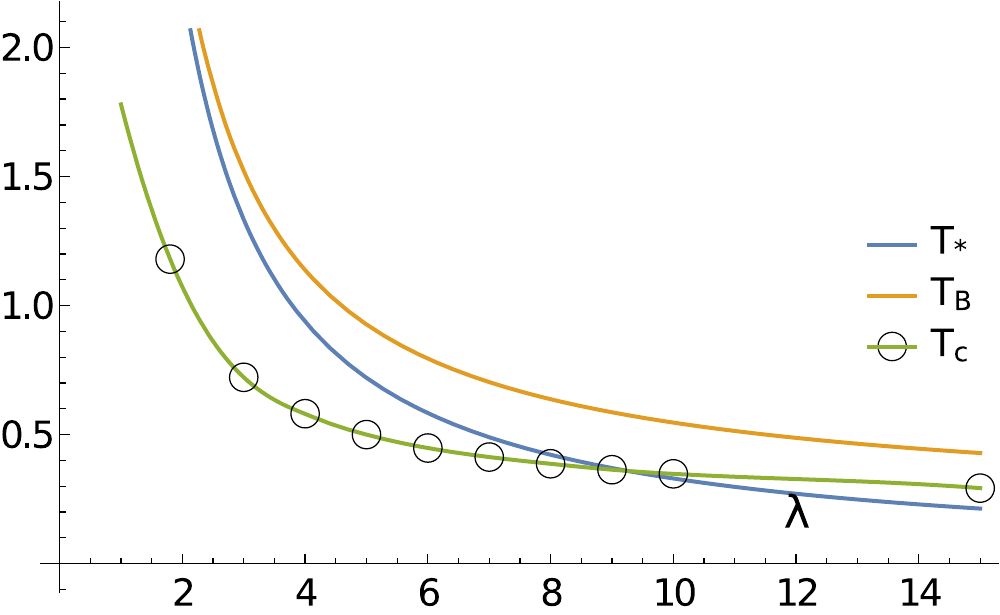}
		\caption{Characteristic temperatures $T_{c}(\lambda)$, $T_{B}(\lambda )$, $T_{*}(\lambda )$ for HCAYF.}
		\label{fig:tstctb}
	\end{figure}

\begin{figure*}
		\begin{subfigure}[b]{0.45\textwidth}
			\centering
		\includegraphics[width=\textwidth]{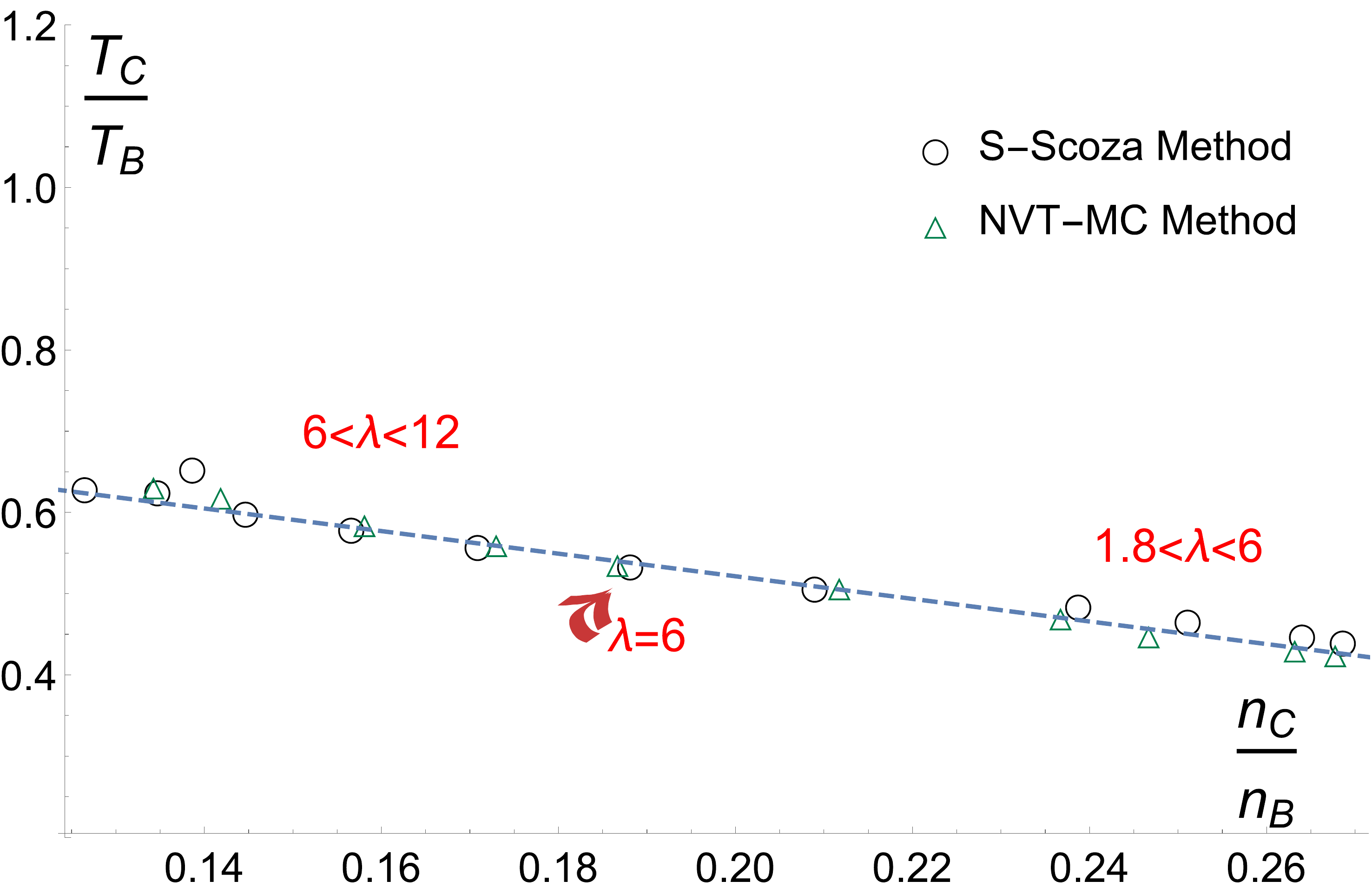}
		\caption{}
		\label{fig:Tc_vs_nC_B}
		\end{subfigure}
\hfil
		\begin{subfigure}[b]{0.45\textwidth}
			\centering		
			\includegraphics[width=\textwidth]{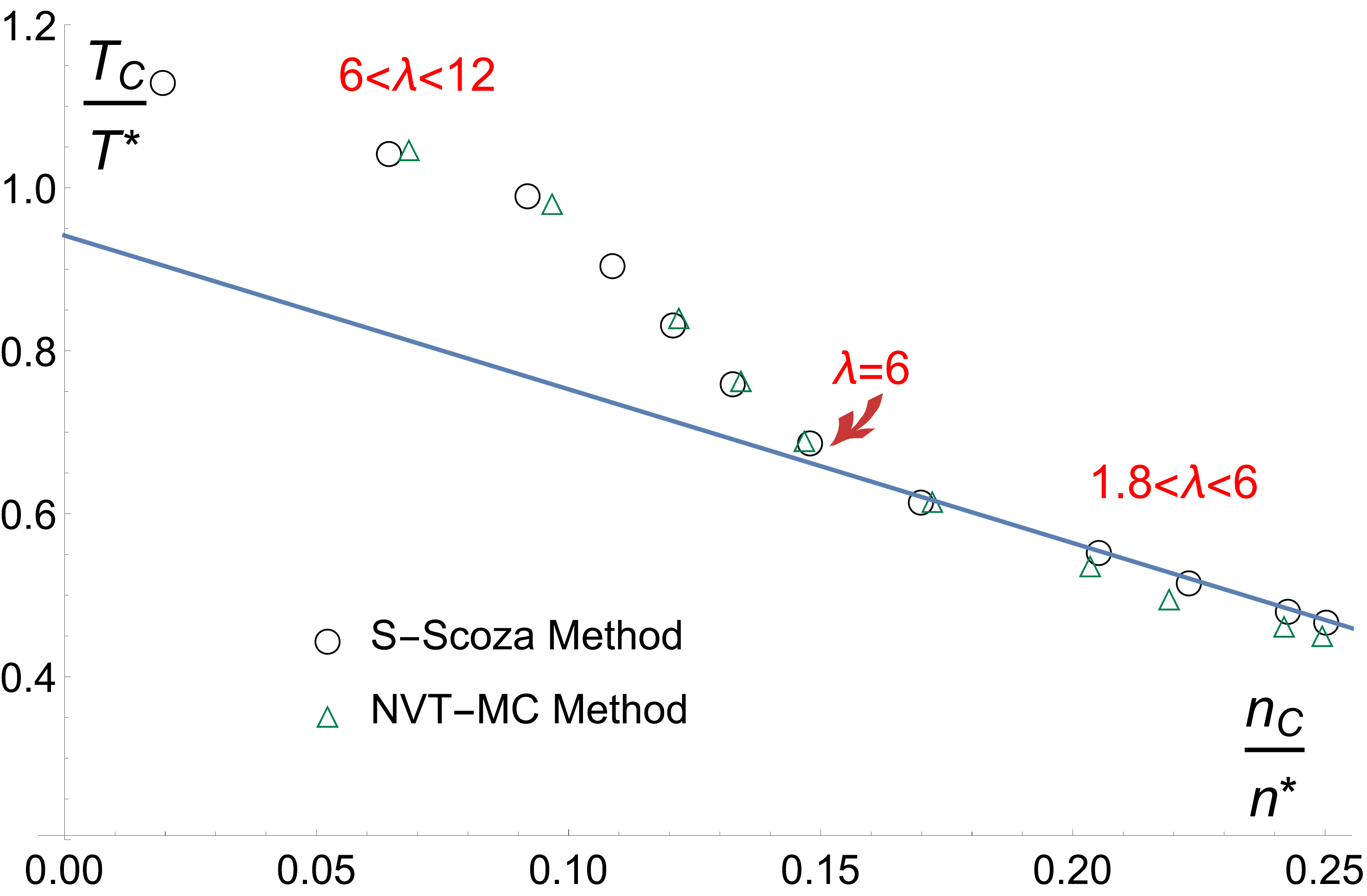}
			\caption{}
			\label{fig:Tc_vs_nC_S}
		\end{subfigure}
	\caption{The line of the critical points for HCAYF normalized by (a) the virial Zeno-parameters $\rho_{B}, T_{B}$ and (b) by the Zeno-element parameters $\rho_{*}, T_{*}$. Critical point data are from \cite{eos_yukawacorrstates_jcp2008}.
	\label{fig:Tc_vs_nC}}
	\end{figure*}

	The temperature parameters $T_{*},\, T_B$ as well as the critical temperature simulation data $T_c(\lambda)$ are shown on Fig.~\ref{fig:tstctb}. As one can see there is $\lambda^*$ so that 
	\begin{equation}\label{eq:lambdat}
	T_{*}\left(\lambda^*\right)=T_c\left(\lambda^*\right)\,\,
	\Rightarrow \,\,\lambda^{*} \approx 9.3 
	\end{equation}
	This value is in good correspondence with the results on the HCAYF liquid-vapor equilibrium \cite{liq_dijkstra_hsyukw_pre2002}.
Thus the temperature parameter $T_{*}(\lambda)$ allows to state that for $\lambda>\lambda^*$ the global isomorphism transformation \eqref{eq:projtransfr} can not be applied. 
Note that $T_{B}>T_{c}$ for all values $\lambda<20$. Therefore virial Zeno-temperature $T_B$ is insensitive to losing stability of liquid phase in contrast to $T_*$.  Additionally, we can analyze the critical point line (see Fig.~\ref{fig:Tc_vs_nC}). The normalization of the critical point locus on the virial parameters $T_{B}\,, \rho_{B}$ produces fairly straight line shown on Fig.~\ref{fig:Tc_vs_nC_B} in the whole data interval $1\le \lambda\le 12$. Using the Zeno-element parameters $T_{*}\,, \rho_{*}$ as the normalization factors reproduces the straight line only for $1\le \lambda \le 6$. But at $\lambda\gtrsim 7$ it starts to deviate from the trend Fig.~\ref{fig:Tc_vs_nC_S} thus the  transformation Eq.~\eqref{eq:projtransfr} fails.

\begin{figure}
		\includegraphics[width=0.8\linewidth]{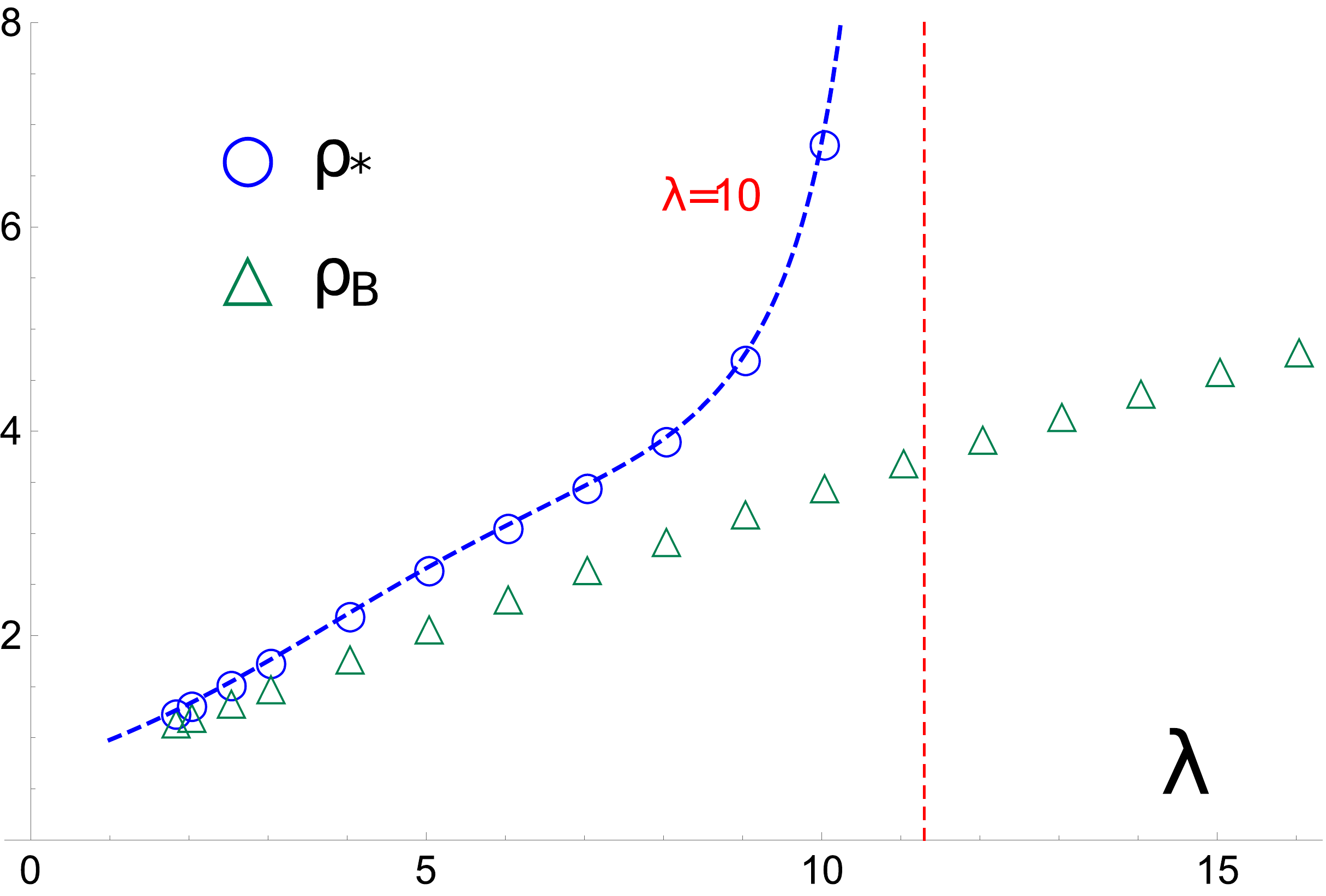}
		\caption{The functions $\rho_{*}(\lambda)$ and $\rho_B(\lambda)$. The Zeno-element density parameter $\rho_{*} \to \infty$ at $\lambda\to 11.3$}
		\label{fig:nSB_vs_Gamma}
	\end{figure}

\begin{figure}
		\includegraphics[width=0.8\linewidth]{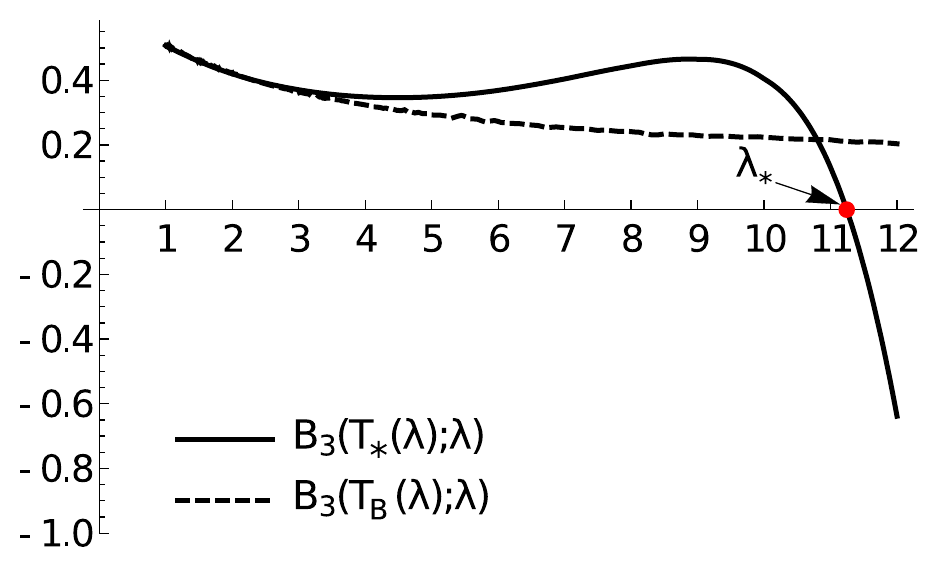}
		\caption{The dependencies $B_3(T_{B}(\lambda);\lambda)$ and $B_3(T_{*}(\lambda);\lambda)$ for HCAYF.}
		\label{fig:b3star}
	\end{figure}

\begin{figure*}
		\begin{subfigure}[b]{0.45\textwidth}
		\includegraphics[width=\textwidth]{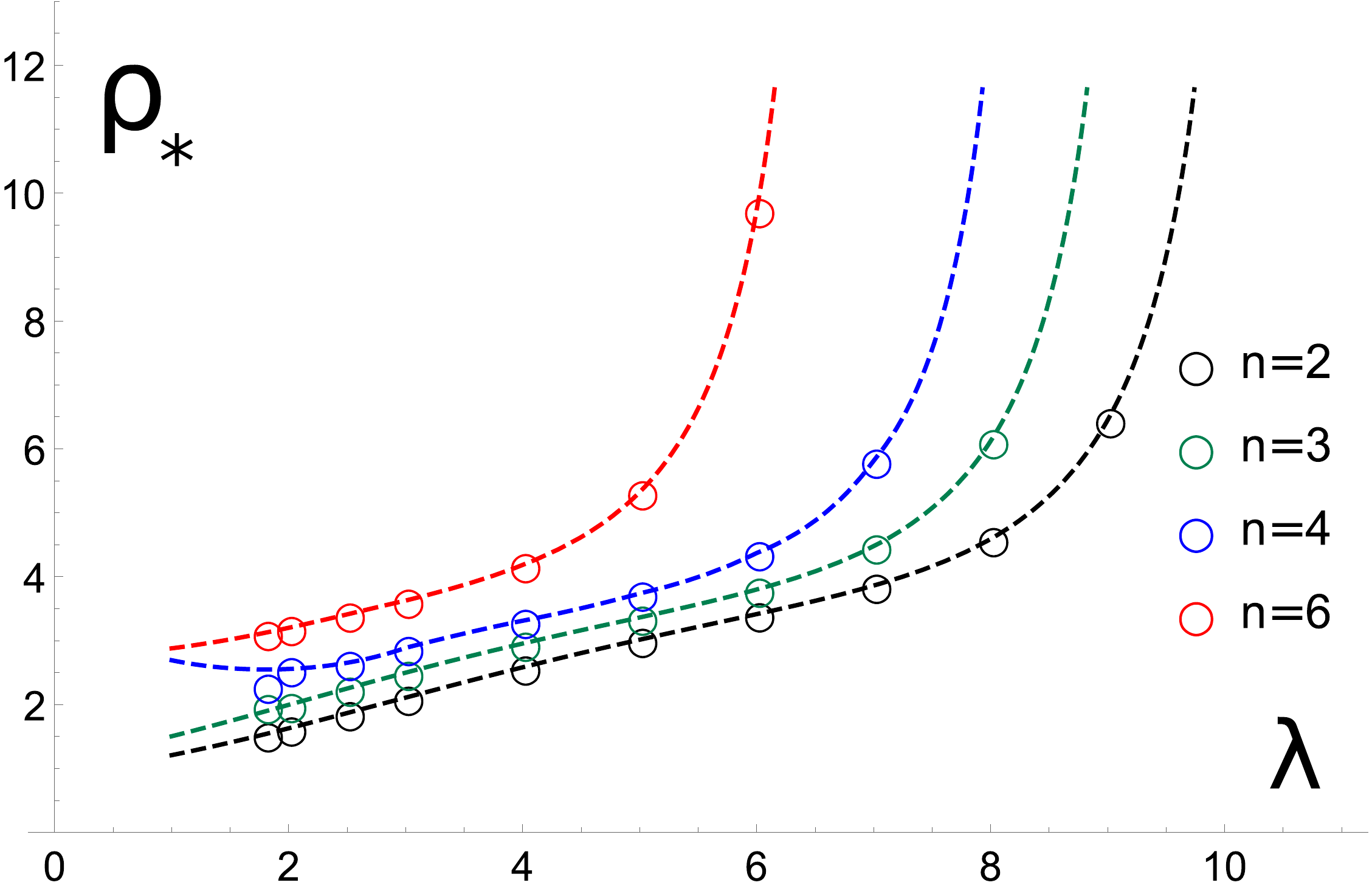}
		\caption{$\rho_{*} (\lambda)$}	
		\end{subfigure}
\hfil
		\begin{subfigure}[b]{0.45\textwidth}		\includegraphics[width=\textwidth]{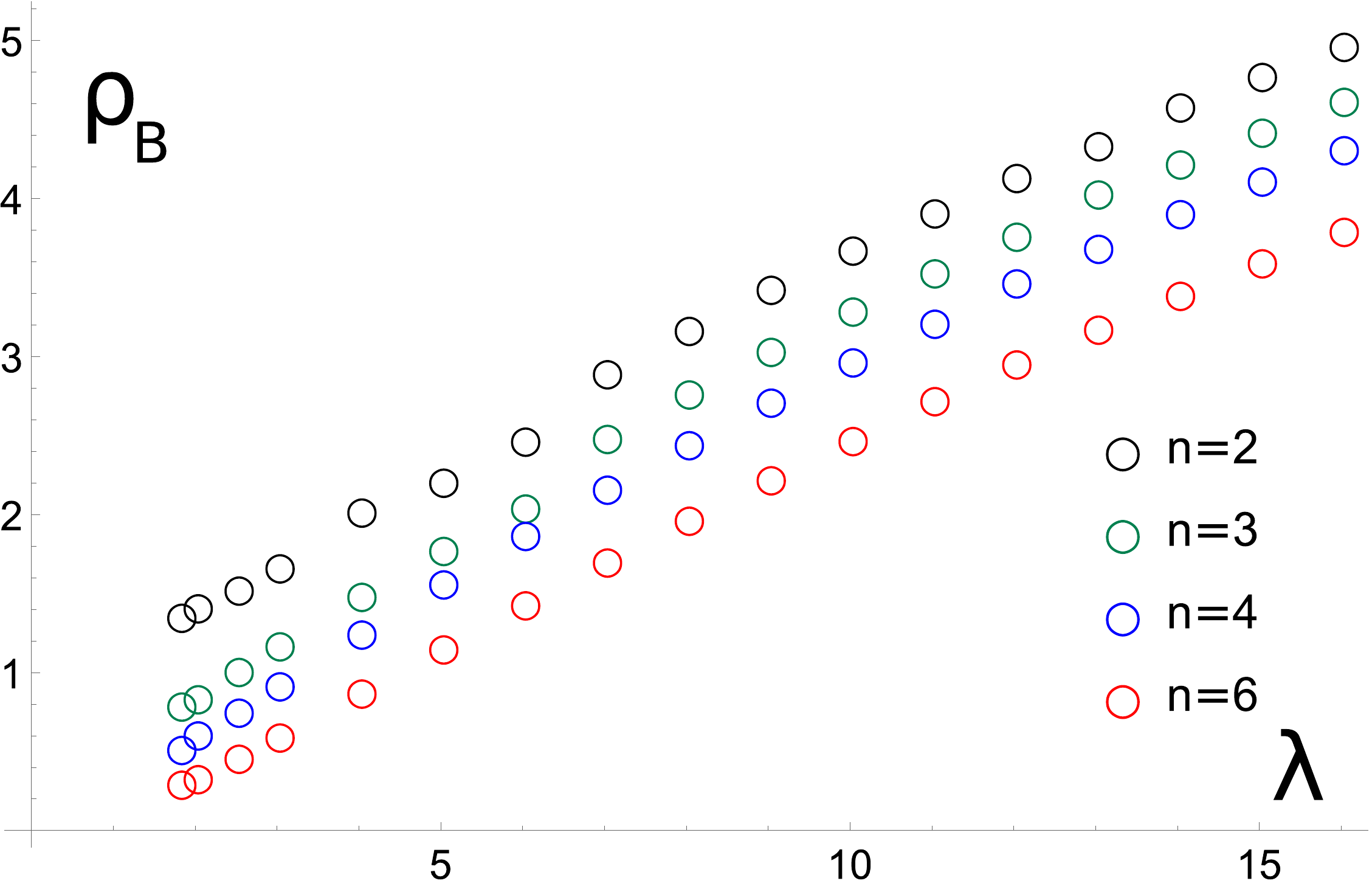}
			\caption{$\rho_B(\lambda)$}
		\end{subfigure}
		\caption{Dependence of the density parameters $\rho_{*}\,, \rho_{B}$ on $\lambda$ for potentials Eq.~\eqref{eq:yukwpot_n}.}
		\label{fig:nSB26_vs_Gamma}
	\end{figure*}

\begin{figure}
		\includegraphics[width=0.8\linewidth]{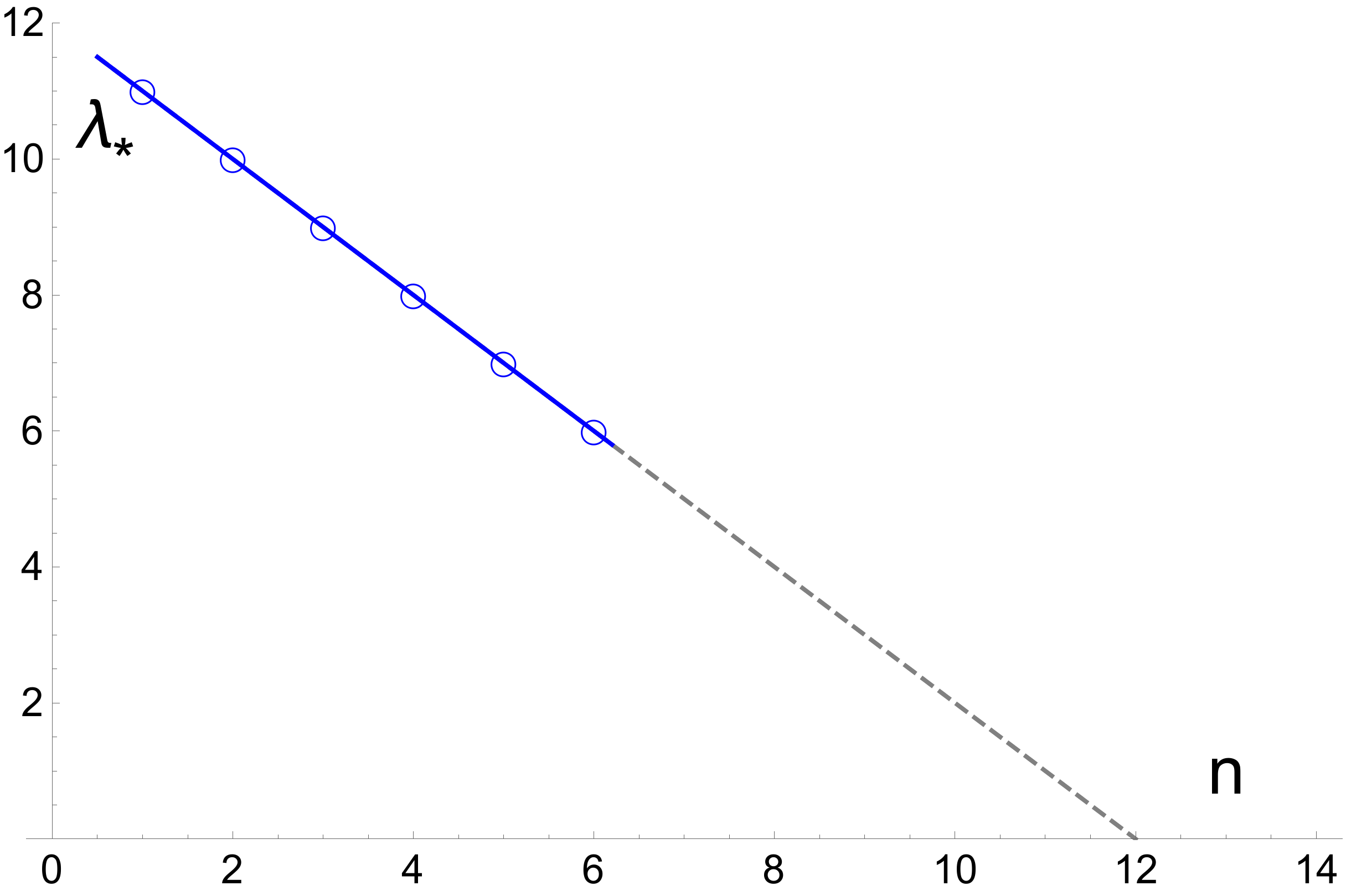}
		\caption{Dependence $\lambda_{*}(n)$ for potentials of the form Eq.~\eqref{eq:yukwpot_n}, dotted line is the linear extrapolation.}
		\label{fig:LambdaX}
	\end{figure}

Now we consider the behavior of the density parameter  $\rho_{*}(\lambda)$ without any reference to the critical point data. The corresponding dependencies  $\rho_{*}(\lambda)$ and $\rho_B(\lambda)$ are on  Fig.~\ref{fig:nSB_vs_Gamma}. There is the divergence of $\rho_{*}$ at $\lambda_{*} \approx 11.3$ because of $B_3(T_{*})=0$ (see Fig.~\ref{fig:b3star}). As has been noted above this signals about the disappearance of the liquid branch of the binodal.
Apparently, such a behavior should be expected also for the screened Sutherland-like potentials 
	\begin{equation}\label{eq:yukwpot_n}
	\Phi_{n}(r) =\begin{cases}
	\infty\,, & \text{if}\quad r< \sigma \\
	-\,\varepsilon\,\left(\frac{\sigma}{r}\right)^n \exp\left(\,-\lambda (r/\sigma -1) \,\right)\,, & \text{if} \quad r\ge \sigma\,,
	\end{cases}
	\end{equation}
	if $\lambda$ is great enough. 
It is quite clear that if $n$ increases the potential becomes more short ranged and therefore the divergence of $\rho_{*}$ happens at lower values of $\lambda$. This is indeed the case. The behavior of $\rho_{*}$ for several values $n\leq6$ is demonstrated on Fig.~\ref{fig:nSB26_vs_Gamma}. 
Therefore we can conclude that the liquid phase becomes unstable for the fluids with potentials \eqref{eq:yukwpot_n} and get the corresponding dependence $\lambda_{*}(n)$ (see Fig.~\ref{fig:LambdaX}). The virial Zeno-parameter $\rho_B$ does not show any peculiarities.
\section{Conclusions}
In this work we have demonstrated the applicability of the global isomorphism approach to the hard-core attractive Yukawa fluids. For interaction range $1\lesssim \lambda \lesssim 7$ the lattice gas isomorphism transformation can be applied to symmetrize the liquid-vapor binodal. Using the available data on liquid-vapor equilibrium of the HCAYF we check the applicability of the projective transformation \eqref{eq:projtransfr} and mapping to the Ising model binodal for such model fluid.
It has been demonstrated that the Zeno-element parameters $T_{*}, \rho_{*}$ appear to be sensitive to the loss of liquid phase stability. Surprisingly, the virial Zeno-line Boyle parameters $T_{B}, \rho_B$ do not show any peculiarities as functions of $\lambda$. This allows to conclude that the global isomorphism transformation Eq.~\eqref{eq:projtransfr} and its parameters $T_{*}, \rho_{*}$ capture nontrivial connection between liquid-vapor equilibrium of continuous fluids and that of the lattice gas models. Technically, the difference between the virial Zeno-line density parameter $\rho_{B}$ and $\rho_{*}$ is that they are determined by different characteristic temperatures $T_{B}$ and $T_{*}=T^{(vdw)}_{B}$ correspondingly. Evidently, $T_{B}$ is determined by the whole interaction potential while $T^{(vdw)}_{B}$ uses its coarse-grained structure in a form of effective hard-core diameter repulsion and the attractive part. Although simplified, such a structure is directly related to that of the lattice gas models of simple Ising-like type. The break of the isomorphism transformation is due to the temperature parameter $T_*$ becomes the null of the third virial coefficient $B_{3}(T_{*}) = 0$ and $\rho_{*}\to \infty$. The reason for using Zeno-element parameters $T_*, \rho_{*}$ as the parameters of the triangle of liquid-gas states is preferable over $T_{B}, \rho_{B}$ is as following. In terms of global isomorphism with corresponding LG (Ising-like) model the low-temperature states $\rho \to 0$ and $\rho \to \rho_*$ inherit the duality of the LG states $x \to 0$ and $x \to 1$. From the physical point of view the state $x\to 1$  can be treated as almost ideal gas of unoccupied cites. Thus the high-density liquid state can be considered as the gas of voids. They interact by the same site-site attractive potential which sets the value of $T_*$. The stability of the liquid phase assumes that such attraction is great enough to ensure that $T_*\gtrsim T_c $ and $\rho_{*} \sim 1$. If the attraction potential is short-ranged and rather weak the critical state can not be achieved and the liquid phase loses its stability. So one can assume that either $\rho_{*} \to 0$ or $\rho_{*} \to \infty$ is expected as the signature of liquid phase instability and breaking of the LG isomorphism. Sure we can not assert that such result is general and serves as a benchmark of the liquid phase stability. Indeed, it is proximity of the triple point to the critical one which governs the stability region of the fluid phase. Obviously, the triple point does not determined by the virial expansion. Nevertheless, the limiting liquid phase stability value $\lambda_{*}$ for the HCAYF is consistent with known results and support simple estimates based on the relation between the loci of triple and critical points \cite{eos_zenomegenpcs_jcp2010}. In this respect it interesting to check the prediction $\lambda_{*}(n)$ for the screened Sutherland potential Eq.~\eqref{eq:yukwpot_n} and obtain the limiting value of the exponent $n_{*}$ at $\lambda_{*} \to 0$. Simple extrapolation of the result on Fig.~\ref{fig:LambdaX} gives the value $n_{*}\approx 12$ which reasonably corresponds to the value $n_{*}\gtrsim 8 $ obtained in \cite{eos_yukawafluid_jcp1994,crit_hardspheres_pre2003,eos_yukawaoz_jcp2008,eos_yukawaliq_jcp2010}.
	
Also the global isomorphism relations  \cite{eos_vliegerthartlekkerkerkerme_jcp2011} between $T_*,\rho_{*}$ and $T_c,\rho_{c}$ fail for the potentials with discontinuous attractive part (square-well potential), or those which produce quite different topology of fluid phase diagram, e.g. with several critical points  \cite{liq_vlieglekkvirialzhou_molsimul2007}. The dependence of the parameter $z$ on the symmetry properties of the potential and contact value of the correlation function at the critical point should be investigated. This will be the subject of the separate paper.
\begin{acknowledgements}
This work was completed due to individual (V.K.) Fulbright Research Grant (IIE ID: PS00245791).
\end{acknowledgements}
%
%
%

\end{document}